\documentclass[]{aa}  
\usepackage{graphicx}
\usepackage{txfonts}
\usepackage{lipsum}
\usepackage[dvipsnames]{xcolor}

\begin{document} 
   \title{Dust dynamics in planet-driven spirals}

   \author{J.A. Sturm
          \inst{1}\thanks{sturm@strw.leidenuniv.nl}
          \and
          G.P. Rosotti\inst{1}
          \and
          C. Dominik\inst{2}
          }

   \institute{
            Leiden Observatory, Leiden University, P.O. Box 9513, NL-2300 RA Leiden, The Netherlands
            \and
            Anton Pannekoek Institute for Astronomy, University of Amsterdam, Science Park 904, 1098 XH Amsterdam, The Netherlands
            }
    \date{Received 14 July 2020 / accepted 14 September 2020}

    \abstract
   {Protoplanetary disks are known to host spiral features that are observed in scattered light, ALMA continuum and more recently in CO gas emission and gas dynamics. It is however unknown if spirals in gas and dust trace the same morphology.}
   {We study the morphology and amplitude of dusty spirals as function of Stokes number and the underlying mechanisms causing a difference from gas spirals. We then construct a model to relate the deviation from Keplerian rotation in the gas to a perturbation in surface density of gas and dust.}
   {We use $\texttt{FARGO-3D}$ with dust implementation to numerically study the spirals, after which the results are interpreted using a semi-analytical model. This model is tested on observational data to predict the perturbation of the spiral in gas dynamics based on the continuum data.}
   {We find that the pitch angle of a spiral does not differ significantly between gas and dust. 
   The amplitude of the dust spiral decreases with Stokes number (St) and starts to fade out at a typical St $>$ 0.1 as the dust becomes decoupled from the gas. The semi-analytical model provides an accurate and fast representation of the spiral in surface density of the dust from the gas. We find a spiral in the TW Hya velocity residual map, never seen before, which is a feature in the vertical velocity and has a kink at the continuum gap, yielding strong evidence for a planet at 99 au.}
   {We built a model that gives an estimate of the underlying dynamics of dust in a spiral, which can serve as a proof of planetary origin of spirals and can be a probe for the Stokes number in the disk.}

    \maketitle

\section{Introduction}
\label{sec:intro}
It has been known for a long time that young stars harbor disks consisting of both dust and gas, that accrete over time onto their host stars.
In the recent years with ALMA and multiple high contrast optical telescopes (e.g. VLT/SPHERE, Gemini/GPI), more and more substructure is found in these protoplanetary disks such as rings, gaps, crescents, clumps and spirals \citep[see for a review e.g.][]{2020arXiv200105007A}. In this paper we focus on spirals. 
Spirals often share similar morphology: two symmetric spiral arms on both sides of the disk, offset by 180$^\mathrm{o}$ in azimuthal direction and often accompanied by radial symmetric gaps close to the spirals.
Spirals are found in scattered light \citep[e.g.][]{2004ApJ...605L..53F,2015A&A...578L...6B,2016AJ....152..222A}, but also in the millimeter continuum \citep[e.g.][]{2016Sci...353.1519P,2018ApJ...869L..43H,2020MNRAS.491.1335R} and more recently in CO gas emission \citep[e.g.][]{2018ApJ...869L..44K,2017ApJ...840...32T} and CO gas dynamics \citep{2019ApJ...884L..56T}.
This list is far from complete and many questions remain open. For example, in some objects, spiral arms are found in gas tracers, but not in the dust continuum, while many spirals in the dust continuum or scattered light do not have a counterpart in the gas.

Some spirals can be explained as the result of the wake created by a massive planet or stellar companion, either located inside or outside the disk, launching at Lindblad-resonances \citep[see e.g.][]{2002MNRAS.330..950O}.
These phenomena are very interesting as they allow us to study the mass of the perturbing planet and properties in the disk that set the shape and morphology of these spirals.
The morphology and pitch angle of these spirals in the gas is well known from both theory \citep{2002ApJ...569..997R,2002MNRAS.330..950O} as well as numerical simulations \citep[see e.g.][]{2015ApJ...809L...5D,2018MNRAS.474L..32J,2018ApJ...859..118B,Veronesi}.
However, some spirals can also be explained as a result of gravitational instability, which happens for high disk masses, specifically when the Toomre $Q$ parameter is lower than 1 \citep{2003MNRAS.339.1025R,2010MNRAS.407..181C,2014MNRAS.444.1919D}.
While the number of spiral arms can vary in the self-gravitating case, for the most massive disks the result in surface density can look very similar to a planet-launched spiral, as in this case they also have two symmetric spiral arms \citep{2015ApJ...812L..32D,2018MNRAS.477.1004H}.
This type of spirals has logarithmic pitch angles, while planet-driven spirals have linear pitch angles, but in most of the sources the spiral signal is too limited to discriminate between the two mechanisms directly, if a companion is not detected \citep{2018ApJ...860L...5F}. Elias 27 is currently the best candidate for gravitational instability as the launching mechanism \citep{2017ApJ...839L..24M}.
One way to discriminate between the two mechanisms can be the gas dynamics, as planet-driven spirals move at the same orbital velocity as the planet, while GI driven spirals move along with the flow.
As it is very time consuming to observe gas dynamics, at the moment of writing only one spiral arm has been observed in the velocity residuals \citep{2019ApJ...884L..56T}.

\citet{2018MNRAS.474L..32J} show that the pitch angle of planetary spirals is higher in the NIR than in the sub-millimeter dust emission, as the two tracers come from different layers in the disk at different sound speeds.
This serves also as a method to test the planetary hypothesis, since the spirals launched by gravitational instability have no significant difference in pitch angle between observational tracers, as GI tends to erase or invert the temperature differences between the midplane and higher layers as a result of shocks in the midplane \citep[e.g.][]{2002ApJ...567L.149B}.
An often made assumption is that the spirals in the ALMA continuum trace the same morphology as the spirals in the midplane gas.
However, the fact that some spirals are visible in the dust but not in the gas or vice versa suggests that this assumption is potentially not always valid.

Both disks and mature planets are observed in great detail, however, the formation of planets is difficult to observe directly.
Dust-growth models and planetesimal formation mechanisms often assume an initial grain size distribution, as we do not have high precision methods to determine the dust grain size in observations.
However, we know that dust particles, pressureless of their own, experience radial drift as they feel a headwind from the gas pressure \citep[see e.g.][]{1972fpp..conf..211W,2002ApJ...581.1344T}.
This means that in general the dynamics of the dust are different from that of the gas. 
By analogy, we can expect changes in the morphology and amplitude of the spirals.
In this work we focus on this problem, studying the differences between spirals in gas and dust and deriving semi-analytical relations that explain the differences. 
Surprisingly, up to now there has been no detailed study on the differences between gas and dust spirals. 
While \citet{Veronesi} performed numerical simulations of dust spirals for different Stokes numbers, they did not interpret them in light of a semi-analytical model as we do in this work.
Understanding the dynamics of planet-launched spirals can give us further insight in the planet formation process and can serve as a probe for the Stokes numbers in disks.  

The paper is structured as follows: we first discuss the methodology in Sect. \ref{sec:methods}. 
We then present the results of the amplitude and morphology differences between gas and dust in Sect. \ref{sec:results_spiral_modelling}, using numerical simulations and the interpretation of these results by deriving a semi-analytical model that constructs the spiral in the dust from the deviations from the Keplerian velocity as a result of the spiral in the gas.
We test our model on ALMA observations in Sect. \ref{sec:results_obsdata}.
We discuss the limitations of our models, alternative explanations and future prospects in the context of spiral arm detections in gas and dust in Sect. \ref{sec:discussion}. We finally draw our conclusions in Sect. \ref{sec:conclusions}.

\section{Methods}\label{sec:methods}
\subsection{Numerical setup}\label{ssec:numerical_setup}
We conduct multi-fluid simulations of the spiral structure formed by a planet potential. 
We evolve the dust and the gas at the same time using the $\texttt{FARGO-3D}$ code \citep{2016ApJS..223...11B}, using the Eulerian dust implementation described in \citet{2016MNRAS.459.2790R}. The implementation uses the semi-implicit integrator introduced by \citet{2015MNRAS.452.3932B}, but applied to a grid-based code rather than to a particle-based code.
$\texttt{FARGO}$ uses the $\texttt{ZEUS}$ numerical algorithm in a co-rotating reference frame.

\subsection{Initial conditions}
We run $\texttt{FARGO-3D}$ in a 2D cylindrical geometry using an evenly spaced grid, ranging from 0.4 to 3 times the radius of the planet's orbit, with a resolution of $N_r$ x $N_\phi$ = 300 x 1000.
The planet is kept on a circular orbit at $r_{\rm pl} = 1$ and we do not allow it to migrate.
We use a locally isothermal model with a constant flaring index, such that the aspect ratio in the disk is given by:
\begin{equation}
    h(r) = \frac{c_\mathrm{s}}{v_\mathrm{k}} = \frac{H}{r} = h_{\rm pl}\, r^{f},
\end{equation}
where $c_\mathrm{s}$ is the sound speed in the medium, $v_\mathrm{k}$ is the Keplerian velocity, $H$ is the pressure scale height, $r$ is the radius in the disk, $h_{\rm pl}$ is the aspect ratio at the position of the planet and $f$ is the flaring index.
We choose $f$ = 0.25 and an aspect ratio at the planet position $h(r_{\rm pl} = 1) = 0.05$, which are typical numbers, but the value of these constants do not affect the results of this paper.\\

The gas surface density in the disk is assumed to follow a single exponent power law
\begin{equation}\label{eq:sigma}
    \Sigma(r) = \Sigma_\mathrm{0} r^{-p},
\end{equation}
where $\Sigma_\mathrm{0}$ is the gas surface density at $r_{\rm pl} = 1$ and $p$ is set to 1.
The value of $\Sigma_\mathrm{0}$ is arbitrary as far as the dynamics is concerned and for this reason we show deviations in the surface density relative to the radial symmetric surface density, so that it can be scaled to all values of $\Sigma_\mathrm{0}$.
For the viscosity, we use the $\alpha$ prescription of \citet{1973A&A....24..337S} and fix $\alpha_\mathrm{\nu}$ = 10$^{-2}$.
We will discuss the impact of a change in $\alpha_\mathrm{\nu}$ on our results in Sect.  \ref{ssec:discussion_dependencealphar}\\

We use 40 dust species having St logarithmically spaced between 10$^{-4}$ and 1, where St is the Stokes number or dimensionless stopping time, defined as
\begin{equation}
    \label{eq:St}
   \mathrm{St} = t_\mathrm{s}\Omega_\mathrm{k} = \frac{a\rho_\mathrm{d}}{\Sigma_\mathrm{g}},
\end{equation}
where $t_\mathrm{s}$ is the stopping time, $\Omega_\mathrm{k}(r)$ is the Keplerian angular velocity at radius r in the disk, $a$ is the size of the dust particles and $\rho_\mathrm{d}$ is the bulk density of the dust, which we assume to be 3.6 $\mathrm{g\, cm}^{-3}$ \citep{1994Icar..111..227L}.
The latter equality is true assuming that the particles are in the Epstein regime which is adequate for protoplanetary disks \citep{2010A&A...513A..79B}.
We choose the maximum St to be well above the maximum grain size in grain-growth models \citep{2012A&A...539A.148B} and below the threshold where feedback from the dust on to the gas can be neglected \citep{2016MNRAS.459.2790R}. 
We choose the minimum St to be small enough such that the dust can be considered closely coupled to the gas as we will show in the next section, so for the purpose of this paper there is no need to take smaller St into account. 
The dust surface density is scaled with respect to the gas surface density using a constant dust-to-gas ratio of 100. Because we ignore the dust feedback on the gas, our simulations can be rescaled to any dust-to-gas ratio as long as it is small.\\

We modeled the spiral using three different planet masses: Earth mass planet $M_{\mathrm{E}}$ = $3\cdot10^{-6}M_\mathrm{\ast}$, Super Earth mass planet $M_\mathrm{SE}$ = $10^{-5} M_\mathrm{\ast}$ and Jupiter mass planet $M_\mathrm{J}$ = $10^{-2} M_\mathrm{\ast}$.
The planet potential is delayed for two orbits and is gradually added to the model, using a taper over 3 orbits.
This prevents high frequency artifacts in the data and makes sure that the spiral sets quickly in time.
We checked that the spiral morphology is in steady state, which is the case after about 10 orbits, so the result after 15 orbits of integration is used in all further analyses.

\subsection{Boundary conditions}
For the gas simulations we used closed boundaries where the radial velocity is mirrored at the boundaries and the azimuthal velocity and surface density are scaled using the closest active cell.
For the dust simulations we used open boundary conditions to avoid the dust piling up at the inner boundary due to radial drift of the larger dust grains.
To make sure that the surface density stays constant over time and to prevent gradual depletion of the large dust grains, the radial velocity at the boundary is set to the radial drift velocity, as given in \citet{2002ApJ...581.1344T}:
\begin{equation}
    v_{r\mathrm{,d}} = \frac{\eta v_\mathrm{k}}{\mathrm{St} + \mathrm{St}^{-1}} \propto \left(\frac{H}{r}\right)^2v_\mathrm{k} \propto r^{2f - 1/2},
\end{equation}
where $\eta$ = $-\left(\frac{H}{r}\right)^2\frac{\mathrm{d}\mathrm{log}(P)}{\mathrm{d}\mathrm{log}(r)}$ with $P$ the gas pressure.
\begin{figure*}
    \centering
    \includegraphics[width=1\textwidth]{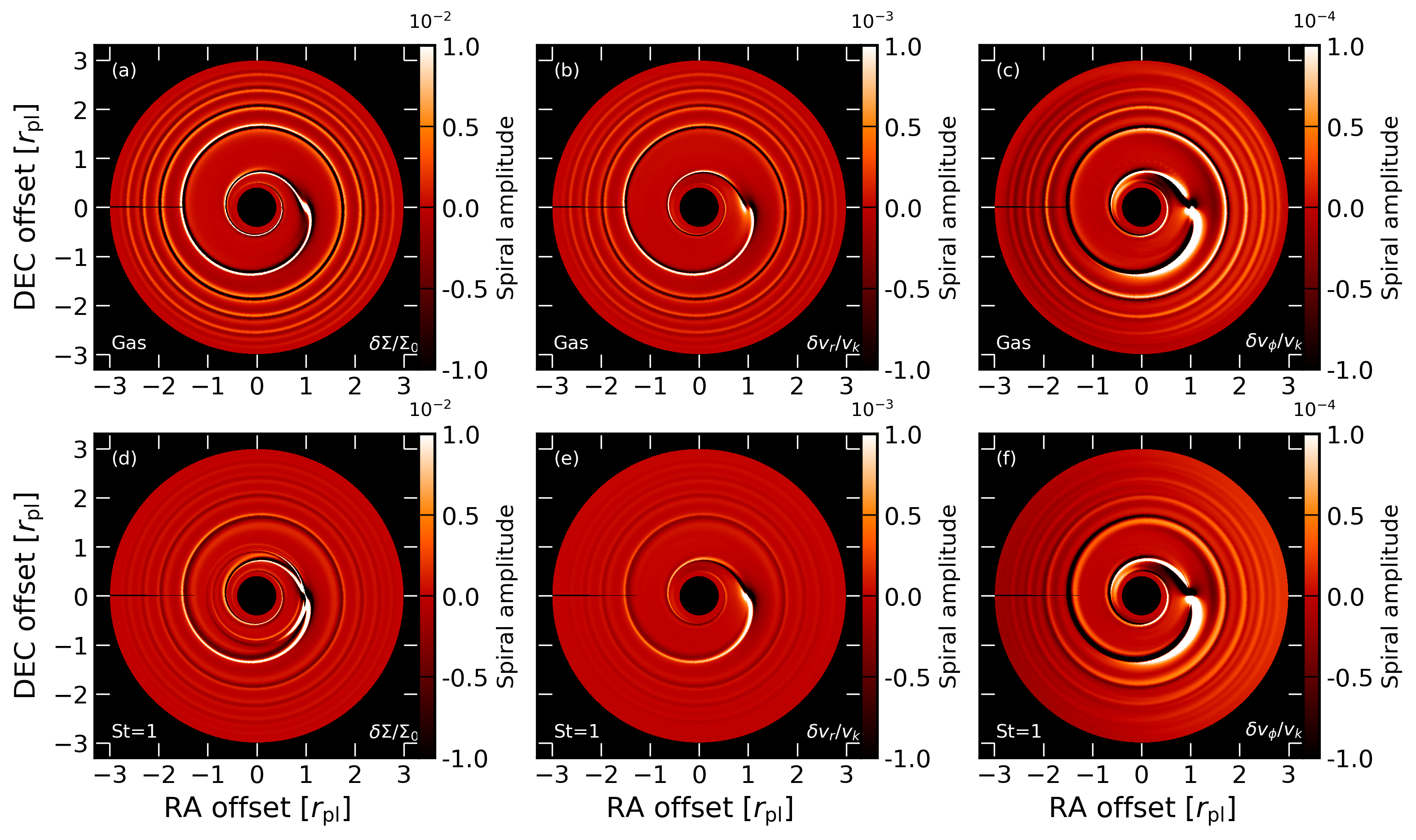}
    \caption{Synthetic $\texttt{FARGO-3D}$ images of spiral arms created by the potential of a super earth mass planet in the gas component (Panels a,b and c) and in the large dust grain component with St = 1 (Panels d,e and f). The images show the perturbation of the spiral normalized to the unperturbed disk, from left to right: the normalized perturbation of the surface density (a and d), the perturbation in radial velocity with respect to the Keplerian velocity (b and e) and the normalized perturbation in azimuthal velocity (c and f).}
    \label{fig:spiral_overview}
\end{figure*}
\begin{figure*}
    \centering
    \includegraphics[width=1\textwidth]{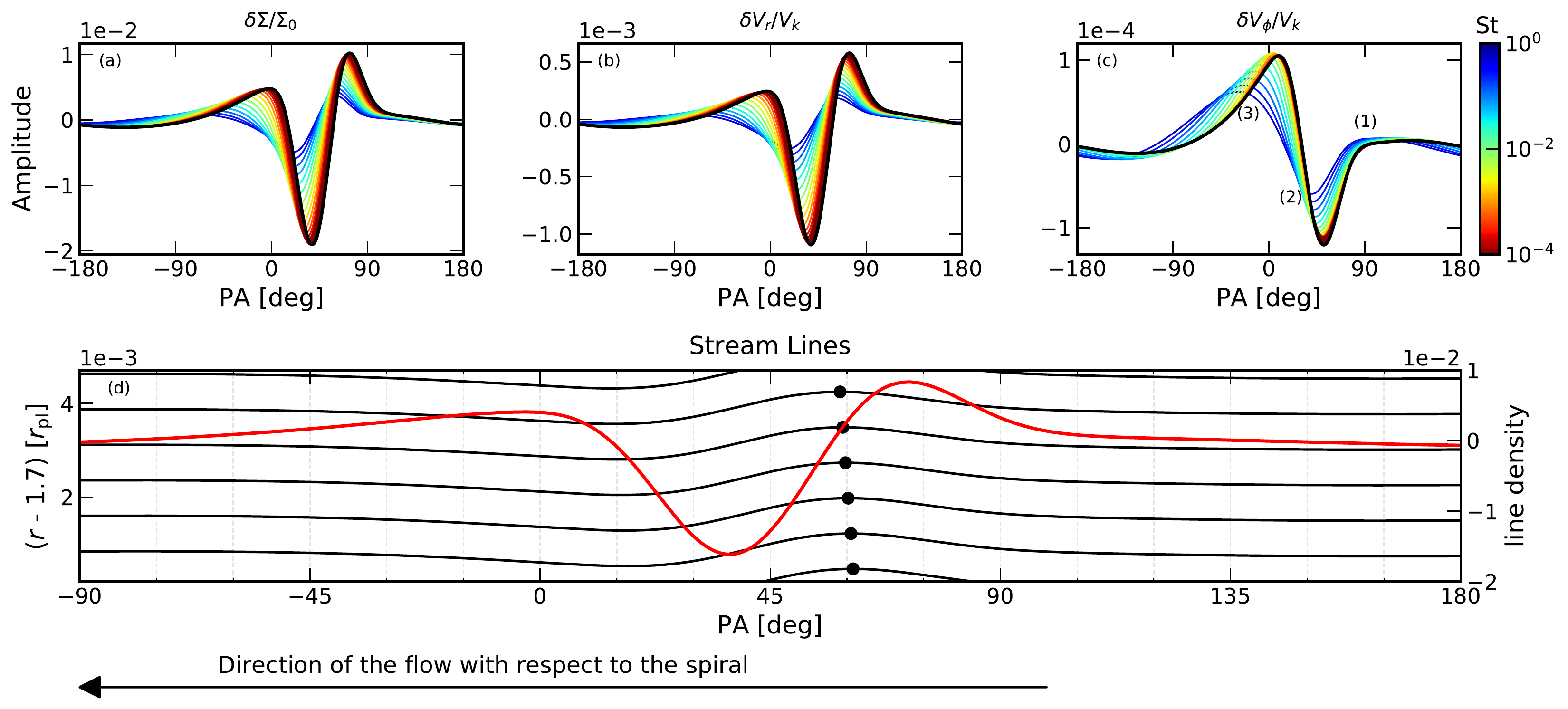}
    \caption{Top: from left to right an azimuthal slab through the normalized perturbation in azimuthal velocity (a), radial velocity (b) and surface density (c) for Stokes numbers 10$^{-4}$ - 1. The perturbation in the gas is indicated in black as a reference.
    Bottom: Streamlines that a gas parcel would follow along an orbit at r = 1.7 r$_{\rm pl}$. The maximum radial deviation is marked with a black dot as a guide for the eye to the effect of the pitch angle on the flow of the particles. The perturbation in line density (see text) is indicated in red, sharing the same shape and amplitude as the perturbation in surface density (panel c).}
    \label{fig:spiral_dynamics}
\end{figure*}

In what follows we find it useful to define a linear pitch angle and we measure it using a linear fit assuming an Archimedean spiral $\phi = ar + b$.
The pitch angle of the spiral can then be written as $\beta = \textrm{tan}^{-1}(a/r)$.
The spiral structure is analyzed at a radius of 1.7 $r_{\rm pl}$ throughout the paper, unless mentioned explicitly. 
This radius is well outside a potential gap opened by the planet, far from any possible boundary artifacts and in the regime where the spiral is approximately Archimedean.
\begin{figure*}
    \centering
    \includegraphics[width=1\textwidth]{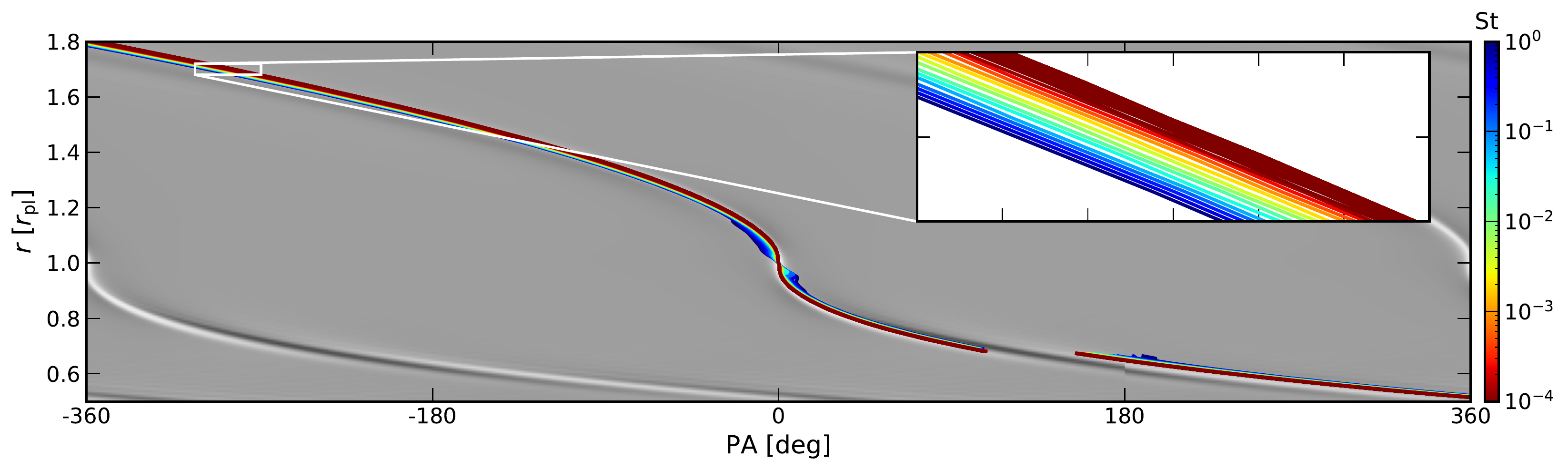}
    \caption{Polar plot of the perturbation in surface density over 2 orbits. The lines indicate the position of the maximum of the different dust components with a zoom of the region around r = 1.7 $r_{\rm pl}$ that is used to determine the change in pitch angle.}
    \label{fig:explanation_pitchangle}
\end{figure*}
\begin{figure*}
    \centering
    \includegraphics[width=1\textwidth]{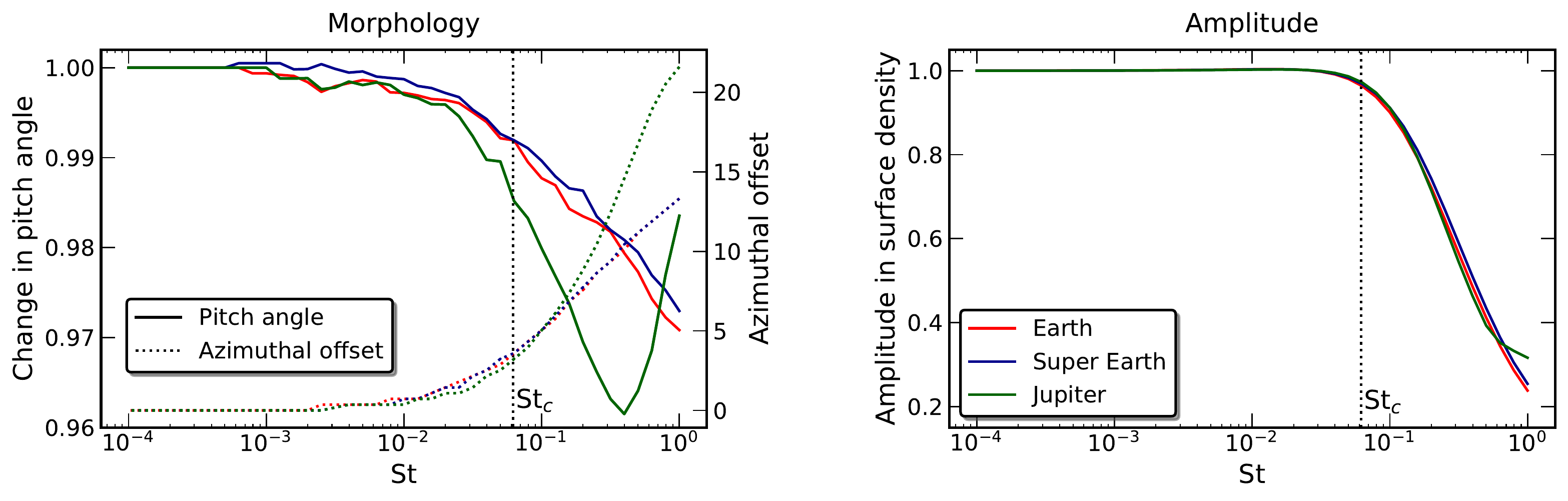}
    \caption{Left panel: The change in pitch angle with respect to the pitch angle of the spiral in the gas, fitted using the function $\phi = ar + b$ between $r$ = {1.6,1.8}r$_{\rm pl}$ as function of Stokes number. The solid lines represent the change in pitch angle relative to the pitch angle in the gas and the dotted lines show the constant azimuthal offset between the spiral in the gas and the dust in degrees.
    Right panel: The amplitude of the spiral versus Stokes number at $r$ = 1.7 $r_\mathrm{pl}$, normalized by the amplitude of the spiral in the gas.
    }
    \label{fig:ang_and_amp_vs_st}
\end{figure*}
\section{Spiral modeling}\label{sec:results_spiral_modelling}
We split the results of our modeling up in three parts. In the first part we have explored the structural changes of the spiral in different dust components and compare this to the spiral in the gas component. 
In the second part we derive semi-analytical relations to derive the density structure of the dust from perturbations in the Keplerian velocity of the gas component and vice versa.
In the third part we combine the relations to explore the physical mechanisms that drive the changes in structure of the spiral in different dust components and show a model that accurately determines the spiral perturbation in the dust surface density from gas dynamics.

\subsection{Comparison between spirals in gas and dust}\label{ssec:results_fargo}
In Fig. \ref{fig:spiral_overview} we show the results of modeling spiral arms created by the potential of a super earth mass planet using $\texttt{FARGO-3D}$ for the gas and for the biggest dust component (St = 1).
We show the amplitude of the spiral in surface density ($\delta \Sigma$), normalized to the surface density of the unperturbed disk ($\Sigma_\mathrm{0}$), and the two components of the velocity perturbation, azimuthal velocity perturbation ($\delta v_\phi$) and radial velocity perturbation ($\delta v_r$), normalized to the Keplerian velocity. 
In terms of the general structure, the spiral perturbation in surface density and the perturbation in radial velocity look largely similar, with the only significant difference that the radial velocity perturbation changes sign at the planets position.
This can be explained using the fact that the spiral moves with the orbital speed of the planet, which means that the gas overtakes the spiral in the inner disk, but the spiral overtakes the flow in the outer disk.
A more detailed description of this can be found in Sect. \ref{ssec:results_model}.
The perturbation in azimuthal velocity has a different structure and is almost an order of magnitude smaller than the perturbation in radial velocity.
The structural difference between the spiral perturbation in the gas and the dust is small, but we see a clear difference in amplitude in all three components.
The results are similar for planets with a different mass, but high mass planets induce a second spiral shifted 180$^\mathrm{o}$ with respect to the first spiral.

In Fig. \ref{fig:spiral_dynamics} we show the azimuthal cross-section of the spiral at a radius of 1.7 $r_{\rm pl}$ in the three components of the perturbation for different St.
The direction of the flow with respect to the spiral is shown from right to left.
The shape of the spiral along azimuth is largely similar in the surface density (panel a) and the radial velocity (panel b), but differs from the azimuthal velocity (panel c).
The peak of the perturbation in the dust components with high Stokes numbers is shifted with respect to the peak of the perturbation in the gas, but the peak always lies on top of the curve of the perturbation in the gas (see Fig. \ref{fig:spiral_dynamics}).
In the azimuthal velocity image we can qualitatively interpret this using the fact that components with high St are no longer closely coupled to the gas, which means that dust particles react delayed to the velocity difference in the spiral; decelerating as long as the gas velocity is lower (from (1) to (2) in the right panel of Fig. \ref{fig:spiral_dynamics}), but never catching up before the gas speeds up as it leaves the spiral (from (2) to (3)).
This results in a similar pattern in the radial velocity and surface density which we will explain in more detail in the next section.

In Sect. \ref{ssec:results_model} (see Eq. \ref{eq:dens_from_angle}) we will go more in detail in analyzing the different components of the velocity perturbation. 
For now, it is useful to consider the stream lines that a gas parcel would follow during an orbit in the vicinity of $r$ = 1.7$r_{\rm pl}$. 
We used a first order Eulerian integrator to determine the stream lines and we have plotted the results in the bottom panel of Fig. \ref{fig:spiral_dynamics} as radial position as function of position angle. 
We also introduce a new quantity, the line density perturbation, shown as a red line. 
This is defined as the density of streamlines in the radial direction. The line density perturbation has the same shape and amplitude as the perturbation in surface density, showing that it is the critical quantity sculpting the spiral. 
The line density is related to the amount of ``squeezing'' in the radial direction, i.e., the radial velocity perturbation. 
For a detailed interpretation of the line density and a toy model explaining this more in detail, see Appendix \ref{app:interpretation_linedensity}.

In Fig. \ref{fig:explanation_pitchangle} we show the location of the dust maxima along azimuthal direction for different St between a radius of 0.5 and 1.8 $r_{\rm pl}$ on top of a polar plot of the surface density perturbation in the gas.
At a radius of 0.7 $r_{\rm pl}$ the maximum of the spiral changes side of the perturbation, causing an azimuthal jump of the maximum.

\subsubsection{Spiral structure as function of St}
\label{sssec:spirals_vs_ST}
\noindent\textbf{Morphology}\\
In the left panel of Fig. \ref{fig:ang_and_amp_vs_st} we show the change in pitch angle of the spiral as a function of Stokes number, representing different particle sizes of the dust, using the results of our simulations.
The linear pitch angle is measured using a fit $\phi = ar + b$ of the position of the maxima along PA between 1.6 $r_{\rm pl}$ < $r$ < 1.8 $r_{\rm pl}$ (see Fig. \ref{fig:explanation_pitchangle}).
There is some difference in the relative change of the pitch angle $\mathrm{tan}^{-1}\mathrm{d}\phi/(r\mathrm{d}r)$ of the spiral for St > St$_\mathrm{c}$ (see next paragraph for a formal definition of St$_\mathrm{c}$), although we find that even for St = 1, the change in d$\phi$/d$r$ is only a few percent. This is less than the constraints we usually can put on the pitch angle due to the uncertainty in disk height and geometry, as well as observational errors. We find instead that there can be a significant azimuthal offset (20 $\deg$ for St = 1) between gas and dust spirals. The offset is higher for Jupiter mass planets.\\

\begin{figure*}
    \centering
    \includegraphics[width=1\textwidth]{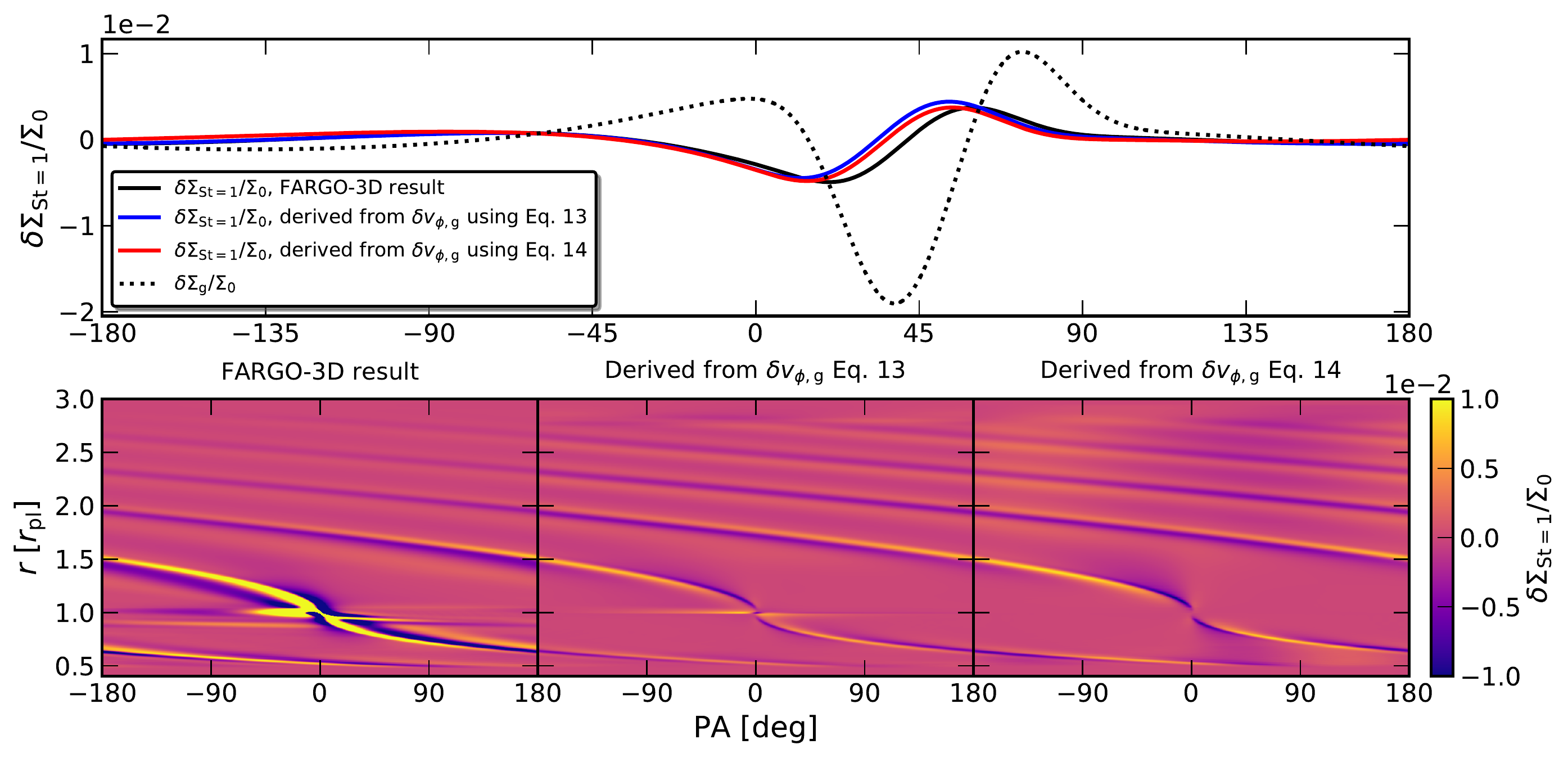}
    \caption{Top: an azimuthal slice through the normalized density perturbation at a radius of 1.7$r_{\rm pl}$ for the dust component with St = 1. The black line shows the time integrated \texttt{FARGO-3D} results, the blue line shows the results directly derived from the perturbation in azimuthal velocity of the gas using the perturbed continuity equation (Eq. \ref{eq:delsigma_from_delvr}), the red line shows the result directly derived from the perturbation in azimuthal velocity of the gas using a direct scaling between $\delta v_r$ and $\delta\Sigma$ and the pitch angle of the spiral (Eq. \ref{eq:dens_from_angle}). The dotted line illustrates the density profile of the gas component for comparison.
    Bottom: The normalized density perturbation in polar coordinates of the result for the St = 1 component. From left to right: the time integrated \texttt{FARGO-3D} result, the results directly derived from the perturbation in azimuthal velocity of the gas and the result directly derived from the perturbation in azimuthal velocity of the gas, using a direct scaling between $\delta v_r$ and $\delta\Sigma$ and the pitch angle of the spiral (Eq. \ref{eq:dens_from_angle}).}
    \label{fig:dustdelsigma_from_gasvphi}
\end{figure*}

\noindent\textbf{Amplitude}\\
In the right panel of Fig. \ref{fig:ang_and_amp_vs_st} we show the maximum amplitude of the spiral at a radius of 1.7 $r_{\rm pl}$ as a function of Stokes number.
The maximum spiral amplitude is for each planet mass normalized to the maximum amplitude of the spiral in the gas component for a better comparison.
We find that the shape of the curve is almost identical for the three planet masses.
Note, however, that the absolute amplitude of the spiral increases if the mass of the planet increases.
The amplitude of the spiral in the surface density of the disk does not change for dust grains with St $\lesssim$ 0.05.
This means that the amplitude of the spiral is constant for almost all dust particle sizes that are available in grain growth models.
To interpret this result we introduce a naive critical Stokes number, determined by the point where the stopping time of the dust is of the same order as the crossing time through the spiral
\begin{equation}
    t_\mathrm{cross} =  \left(\Delta \varphi + \Omega_\mathrm{p}t_\mathrm{cross}\right) t_\mathrm{dyn} = \frac{\Delta \varphi t_\mathrm{dyn}}{1 - \Omega_\mathrm{p}t_\mathrm{dyn}},
\end{equation}
where $\Omega_p$ is the angular velocity of the planet and hence of the spiral, $t_\mathrm{dyn}$ is the dynamical time and $\Delta \varphi$ is the width of the spiral with respect to a whole revolution. The second term arises due to the fact that the spiral rotates slightly during crossing over. Using Eq. \ref{eq:St} the critical Stokes number can be estimated as
\begin{equation}
    \label{eq:Stc}
    \mathrm{St}_\mathrm{c} = \frac{\Delta \varphi}{(1 - \Omega_\mathrm{pl}t_\mathrm{dyn})}.
\end{equation}
Estimating the width of the spiral using a Gaussian fit, we get St$_\mathrm{c}$ = 0.057, 0.061, 0.063 for $M_\mathrm{E}$, $M_\mathrm{SE}$ and $M_\mathrm{J}$ respectively.
The mean of these three numbers is shown in Fig. \ref{fig:ang_and_amp_vs_st} as a vertical dotted line, which matches accurately with the turning point of the graph.
The critical Stokes number depends on $r$, but for $r$ far enough from the planets position, the shape of the curve and the critical Stokes number are very similar as the result shown (see Appendix \ref{app:Stc_vs_r}).\\

\subsection{Semi-analytical description spiral shape}\label{ssec:results_model}
Using analytical relations we are able to determine the perturbation in surface density in the dust from the perturbation in azimuthal velocity in the gas alone.
Three substeps are needed to interpret the results in Fig. \ref{fig:ang_and_amp_vs_st} and understand the dynamics of dust while crossing the spiral: 
\begin{itemize}
    \item The relation between the perturbation in azimuthal velocity and radial velocity
    \item How the perturbation in surface density arises from the perturbations in velocity 
    \item How the perturbations in the velocity of the dust are related to the perturbations in the velocity of the gas.
\end{itemize}

\subsubsection{From $\delta v_\phi$ to $\delta v_r$}
We can write the surface density as an azimuthal symmetric part and a small perturbation $\Sigma = \Sigma_\mathrm{0} $+$ \delta\Sigma$ and the corresponding velocity field as $\vec{v}$ = $\vec{v}_\mathrm{0}$ + $\delta\vec{v}$; $v_{r} = v_{r\mathrm{,0}} + \delta v_r$ and $v_{\phi} = v_{\phi\mathrm{,0}} + \delta v_\phi$, where $v_r$ is the radial component of the velocity and $v_\phi$ is the azimuthal component of the velocity.
Assuming that $\delta v_\phi << v_{\phi\mathrm{,0}}$ and using that a perturbation in Keplerian velocity results in an immediate change of angular momentum, the effect of the perturbation in $v_\phi$ can be explained as spiraling in or outwards on circular orbits:
\begin{equation}\label{eq:vphi}
    v_\phi \simeq r\Omega_\mathrm{k} = \sqrt{\frac{GM}{r}}
\end{equation}
\begin{equation}\label{eq:vr}
    v_r = \frac{\mathrm{d}r}{\mathrm{d}t} = \frac{\mathrm{d}}{\mathrm{d}t}\frac{GM}{v_\phi^2} = GM\frac{\mathrm{d}\phi}{\mathrm{d}t}\frac{\mathrm{d} v_\phi^{-2}}{\mathrm{d}\phi} = GM\frac{v_\phi}{r}\left(\frac{-2}{v_\phi^3r}\frac{\mathrm{d} v_\phi}{\mathrm{d}\phi}\right).
\end{equation}
Assuming $v_{r\mathrm{,0}} = 0$ and using that $\partial_\phi v_{\phi\mathrm{,0}} = 0$, the perturbation in $v_r$ is related to the derivative of $v_\phi$:
\begin{equation}\label{eq:delvr_from_delvphi}
    \delta v_r = -2 \mathrm{sgn}(r - r_{\rm pl}) \partial_\phi{\delta v_\phi},
\end{equation}
where the $\mathrm{sgn}(r - r_{\rm pl})$ term arises from the fact that the spiral moves with the same speed as the planet, which means that outside the planet orbit the spiral is moving faster than the gas, but inside the planet orbit the gas will overtake the spiral, causing a sign flip in the derivative.

\subsubsection{From $\delta v$ to $\delta\Sigma$}\label{sssec:deltav_to_deltasigma}
When the density perturbation is small compared to the average density, we can determine a relation between the density perturbation and the perturbation in gas dynamics caused by the planet potential.
We know that both $\Sigma$ and $\Sigma_\mathrm{0}$ have to obey the continuity equation, resulting in the following set of equations:
\begin{equation}\label{eq:continuity_equation}
    \partial\mathrm{_t} \Sigma_\mathrm{0} + \vec{v_\mathrm{0}}\cdot (\nabla \Sigma_\mathrm{0}) + \Sigma_\mathrm{0}(\nabla\cdot\vec{v_\mathrm{0}}) = 0
\end{equation}
\begin{equation}
    \partial_\mathrm{t}(\Sigma_\mathrm{0} + \delta \Sigma) + (\vec{v}_\mathrm{0} + \delta \vec{v})\cdot (\nabla (\Sigma + \delta\Sigma)) + (\Sigma + \delta \Sigma)(\nabla\cdot(\vec{v}_\mathrm{0} + \delta \vec{v})) = 0.    
\end{equation}
Neglecting quadratic terms in perturbation and assuming that $v_{r\mathrm{,0}}$ = 0, we find
\begin{equation}
    \partial_\phi\delta\Sigma = -\frac{r\Sigma_\mathrm{0}\delta v_r\left(\frac{\partial_r \Sigma_\mathrm{0}}{\Sigma_\mathrm{0}} + \frac{\partial_r \delta v_r}{\delta v_r} + \frac{1}{r}\right) + \Sigma_\mathrm{0}\partial_\phi (\delta v_\phi)}{v_{\phi\mathrm{,0}}}.
\end{equation}
This can be simplified using Eq. \ref{eq:sigma} and Eq. \ref{eq:delvr_from_delvphi} to 
\begin{equation}\label{eq:delsigma_from_delvr}
    \partial_\phi\delta\Sigma = -\frac{r\Sigma_\mathrm{0}\delta v_r}{v_{\phi\mathrm{,0}}}\left(\frac{\partial_r \delta v_r}{\delta v_r} + \frac{1}{2r} - r^{-p}\right),
\end{equation}
where the first term is far dominant in all cases, as we will show in Sect. \ref{ssec:results_combining_everything}.\\

The perturbation in surface density has the same structure as the perturbation in radial velocity, as we illustrate in Fig. \ref{fig:spiral_dynamics}.
This can be interpreted using that the spiral perturbation can be described as a 1D line in \{r,$\phi$\}, as long as the azimuthal shape of the spiral does not change significantly in a radial cut through the spiral.
The relation between taking the radial derivative and the azimuthal derivative is then given by the pitch angle of the spiral, $\mathrm{d}\phi/\mathrm{d}r = a$ with a the linear pitch angle.
Using this assumption Eq. \ref{eq:delsigma_from_delvr} reduces to a simple scalar relation between $\delta\Sigma$ and $\delta v_r$:
\begin{equation}\label{eq:dens_from_angle}
    \delta\Sigma = -\mathrm{sgn}(r - r_{\rm pl})\Sigma_\mathrm{0}r\frac{\delta v_r}{v_{\phi\mathrm{,0}}}\frac{\mathrm{d}\phi(r)_{\rm sp}}{\mathrm{d}r}.
\end{equation}
This result can be interpreted using the bottom panel of Fig. \ref{fig:spiral_dynamics}, where we show stream lines at a few orbits close together that a gas parcel would follow during an orbit around the central object.
Since the positional changes due to the spiral potential are tiny, the azimuthal shape of the spiral can be interpreted as being locally independent of orbit radius.
The only change between stream lines that are close together, is the small azimuthal offset associated with the pitch angle of the spiral. 
This results in a vertical squeezing that is related to the amount of radial change i.e. the radial velocity.
A simple sketch for a detailed explanation of this relation is given in Appendix \ref{app:interpretation_linedensity}.

\subsubsection{From gas to dust}
The velocity perturbation of the dust can be derived from the dynamics of the gas by solving the azimuthal equation of motion for the dust.
The only force acting on the dust is the drag force caused by the gas moving at slightly sub-Keplerian velocity due to the gas pressure.
The equation of motion for the azimuthal velocity of the dust is hence given by:
\begin{equation}
    v_{\phi\mathrm{,d}}\frac{\mathrm{d}v_{\phi\mathrm{,d}}}{\mathrm{d}\phi} = -\frac{r}{t_{\rm fric}}(v_{\phi\mathrm{,d}} - v_{\phi\mathrm{,g}}),
\end{equation}
where the subscript d is the dust component and g is the gas component.
Rewriting this in terms of Stokes number using Eq. \ref{eq:St} we can write the perturbation in azimuthal velocity of the dust as the first order non-linear ordinary differential equation:
\begin{equation}\label{eq:from_gas_to_dust}
    \frac{\mathrm{d}\delta v_{\phi\mathrm{,d}}}{\mathrm{d}\phi} = \frac{r}{\mathrm{St}\Omega_\mathrm{k}}\left(\frac{v_{\phi\mathrm{,g}}(\phi)}{\delta  v_{\phi\mathrm{,d}} + \Omega_\mathrm{k}r} - 1\right).
\end{equation}
We approximated this differential equation with the standard Runge-Kutta method using the $\texttt{integrate.solve\_ivp}$ function in $\texttt{scipy}$ \citep{2020SciPy-NMeth}.
We integrate for three orbits in steps of 0.01 radians, using an initial value for the perturbation of 0 at all radii.

\subsection{Combining everything}\label{ssec:results_combining_everything}
Combining the equations in Sect. \ref{ssec:results_model} makes it possible to determine the shape and amplitude of the spiral in dust surface density directly from the azimuthal velocity perturbation in the gas.
This model has three steps:
\begin{enumerate}
    \item Determine the perturbation in azimuthal velocity of the \textrm{dust} from the azimuthal velocity of the \textrm{gas} using Eq. \ref{eq:from_gas_to_dust}.
    \item Determine the perturbation in \textrm{radial} velocity of the dust from the \textrm{azimuthal} velocity of the dust using Eq. \ref{eq:delvr_from_delvphi}
    \item Determine the perturbation in \textrm{surface density} of the dust from the perturbation in \textrm{radial velocity} of the dust. This can be done using two different approaches.
    \begin{enumerate}
        \item Using the full perturbed continuity equation (Eq. \ref{eq:delsigma_from_delvr})
        \item Using the perturbed continuity equation and numerical derivation of the spiral position to determine the spiral pitch angle (d$\phi$/d$r$) as function of radius to relate the azimuthal derivative with the radial derivative (Eq. \ref{eq:dens_from_angle}).
    \end{enumerate}
\end{enumerate}

The accuracy of the different substeps compared to the $\texttt{FARGO-3D}$ output are given in Appendix \ref{app:substeps}.
In Fig. \ref{fig:dustdelsigma_from_gasvphi} we show the end results in the most extreme case, St = 1.
In the top panel of the figure we show the normalized amplitude of the spiral throughout the whole disk as determined using the aforementioned procedure and on the bottom panel of the figure we show an azimuthal slice through all results as a comparison.
The shape of the perturbation matches accurately with the predicted result from $\texttt{FARGO-3D}$ for $r$ > 1.5 $r_{\rm pl}$.
We find that the two approaches to determine the surface density from the perturbation in radial velocity give almost identical results, justifying the assumption that the perturbation in radial velocity sets the perturbation in surface density as explained in Sect. \ref{sssec:deltav_to_deltasigma}.
In general, the surface density perturbation is more accurately recovered closer to the planet for smaller Stokes numbers.
When Stokes numbers are low enough to consider the dust closely coupled to the gas, the surface density does not change with respect to the gas and the right hand side of Eq. \ref{eq:from_gas_to_dust} approaches zero, not adding any additional uncertainties to the final result.

Combining the equations in Sect. \ref{ssec:results_model} gives a full understanding of the shape of the curves amplitude and pitch angle as function of Stokes number (Fig. \ref{fig:ang_and_amp_vs_st}).
Using Fig. \ref{fig:spiral_dynamics}c we find that dust particles feel the change in the amount of drag when the azimuthal velocity of the gas changes (point (1) as indicated in the figure).
Only very large dust particles with high Stokes number have a reaction time that is longer than the crossing time through the spiral (Eq. \ref{eq:Stc}), so they accelerate and decelerate less than the gas. 
Since the dust particles leave the spiral before they are able to catch up (point (2) and (3) as indicated in Fig. \ref{fig:spiral_dynamics}), the amplitude of the spiral in azimuthal velocity decreases towards larger Stokes number.
A smaller perturbation in azimuthal velocity with a damped acceleration and deceleration, means a smaller perturbation in radial velocity (Eq. \ref{eq:delvr_from_delvphi}), hence in surface density (Eq. \ref{eq:dens_from_angle}).
A consequence of this mechanism is that the peak in azimuthal velocity of the dust is always the intersection point with the azimuthal velocity of the gas, as the dust starts to decelerate after that point (see Fig. \ref{fig:spiral_dynamics}).
This explains the small change in pitch angle of the spiral for larger St (Fig. \ref{fig:ang_and_amp_vs_st}).
Since the spiral decreases in amplitude over radius and broadens in azimuthal shape, the position of the spiral in the dust is not changed by only a constant but the pitch angle changes as well.\\

\section{Testing on observational data} \label{sec:results_obsdata}
\subsection{Source selection}
\label{ss:potential sources}
Our models show that high SNR spiral structures in the gas dynamics, together with a spiral seen in the dust continuum surface density, can put constraints on grain size comparing the observed amplitude of the spiral in the dust with that in the gas.
There are multiple candidate sources that are known to have spirals in the continuum or the gas.
However, to give an accurate estimate of the surface density and to derive a velocity residual map, we need data with high spatial and velocity resolution.
Out of the recent high-resolution DSHARP survey of 20 disks, 3 were found to harbor spirals: Elias 27, WaOph 6 and IM Lup  \citep{2018ApJ...869L..43H}.
Elias 27 harbors the highest contrast spiral arms of the survey, but gravitational instability is a convincing explanation for this source (\citealt{2015MNRAS.453.1768P,2017ApJ...839L..24M,2018ApJ...860L...5F}, although see also \citealt{2018MNRAS.477.1004H}) which makes it unusable for this purpose, as further discussed in Sect. \ref{ssec:discussion_spiral_origin}.
Furthermore, Elias 27 and WaOph 6 suffer from cloud and outflow contamination in the CO gas \citep{2018ApJ...869L..41A}, which makes it impossible to use the kinematic data published so far.
IM Lup disk has two high contrast spiral arms in the mm continuum, and no cloud contamination \citep{2018ApJ...869L..41A}, so is an ideal candidate to analyze.
Unfortunately, no spiral signal is detected in the gas for the IM Lup disk, but in paragraph \ref{ss:imlup} we will show that we can make a prediction on what the spiral looks in the two velocity components.
Other sources are considered, by surveying the literature.
MWC 758 is a promising source \citep{2018ApJ...860..124D,2018ApJ...853..162B}, but the observations are currently too limited to be used.
HD100453 has a spiral detected in scattered light, continuum and CO gas \citep{2017A&A...597A..42B,2020MNRAS.491.1335R}, but the spiral seen in the gas is beyond the size of the continuum disk and the inner region of the disk has a noisy velocity map, potentially due to a warp in the inner disk, which makes the source not useful in our analysis.
TW Hya harbors the only spirals in gas dynamics, potentially in the azimuthal velocity \citep{2019ApJ...884L..56T}.
The spirals are detected in the CO gas emission as well, but no spiral signal is present in the dust continuum images.
In the next subsection, we will analyze the TW Hya disk further and show that the observed spiral signal in the dynamics is interesting for further study, but not useful to test our model. 

\subsection{TW Hya disk}
\label{ss:twhya}

\begin{figure}[ht]
    \centering
    \includegraphics[width=1\linewidth]{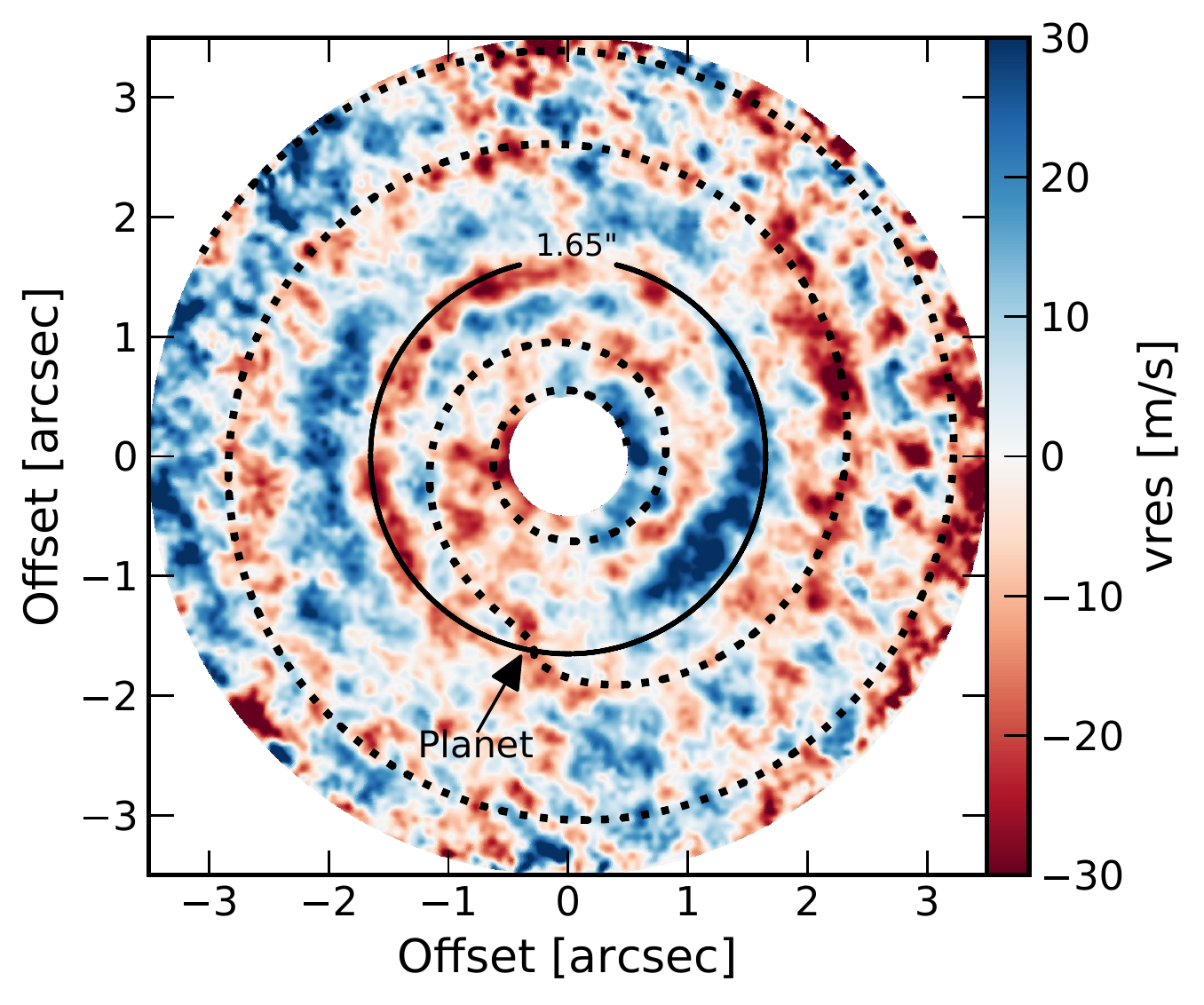}
    \caption{Map of the velocity residual as in \citet{2019ApJ...884L..56T} for the disk around TW Hya, but with flipped colorbar to highlight the blueshifted emission (now seen in red). The dotted line serves as a guide for the eye and is determined using the theoretical shape of a spiral launched by a planet (Eq. \ref{eq:theoretical_spiral}). The solid line represents the radius of the potential planets orbit which coincides with the blue and red shifted circular lobes on either side that indicate sub-Keplerian rotation, probably due to a lower opacity inside the gap, so we probe deeper layers in the disk.}
    \label{fig:twhya}
\end{figure}

TW Hya, located at a distance of 60.1 pc, hosts a well studied protoplanetary disk for which a potential spiral in azimuthal velocity is suggested by \citet{2019ApJ...884L..56T}.
However, using the exact same data set we found a different spiral than originally suggested in the paper, which radically changes the interpretation of the velocity deviations.
For a detailed description of the observations, calibration and further processing of the data, see the original paper \citep{2019ApJ...884L..56T}. 
In Fig. \ref{fig:twhya} we show the exact same data as in Fig. 1 of the original paper, but to a further extent, and derotated such that the semi-minor axis of the disk is vertical.
The colorbar that was used emphasizes the residual positive velocity, but the residual negative velocity can not be neglected in this case.
Flipping the colorbar, now emphasizing the blue-shifted emission, we find a different spiral that has comparable morphology as one of the spirals found in the CO gas emission.
The spiral is constant in amplitude along azimuth and appears to have a kink at the position of the gap in the continuum emission at $r$ = 1.65" and PA = 160$^\mathrm{o}$.
The shape of the spiral is consistent with the theoretical shape given by the wake equation of \citet{2002ApJ...569..997R}, shown as a dotted line in Fig. \ref{fig:twhya}:

\begin{equation}
\label{eq:theoretical_spiral}
\begin{array}{l}
\phi(r) = \phi_{\mathrm{pl}} - \frac{\mathrm{sgn}\left(r - r_{\mathrm{pl}}\right)}{h_{\mathrm{pl}}}\left(\frac{r}{r_{\mathrm{pl}}}\right)^{1 + \beta}\left\{\frac{1}{1 + \beta} - \frac{1}{1 - \alpha + \beta}\left(\frac{r}{r_{\mathrm{pl}}}\right)^{-\alpha}\right\} \\
\quad + \frac{\mathrm{sgn}\left(r - r_{\mathrm{pl}}\right)}{h_{\mathrm{pl}}}\left(\frac{1}{1 + \beta} - \frac{1}{1 - \alpha + \beta}\right)
\end{array},
\end{equation}
where $\alpha$ is the power exponent of the rotation angular frequency ($\Omega(r)\propto r^{-\alpha}$), $\beta$ is the power exponent of the radial distribution of the sound speed ($cs \propto r^{-\beta}$).
We set $\alpha$ to the typical value 1.5 and $\beta$ to 0.3 as determined by \citet{2016A&A...592A..83K} to model the disk.
Using a fit by eye, we find $h_\mathrm{pl}$ = 0.06 as a best fit, which is close to the value used in \citet{2016A&A...592A..83K} to model the disk $(h/r)_\mathrm{pl} \sim$ 0.1.
The fact that we see the whole spiral as blueshifted relative to the disk means that the spiral is a perturbation in vertical velocity, since both the radial velocity as well as the azimuthal velocity perturbations change sign along either the major or minor axis of the disk \citep[see][for a schematic of the different velocity components in a velocity residual map]{2019ApJ...884L..56T}.
These two clues provide potential evidence that a planet is carving out the gap, inducing a spiral that stirs the material up to larger heights in the disk.
The spiral found in the original paper could be part of the sub-Keplerian rotation that we observe due to the fact that we look deeper inside the disk at the location of the gap cleared by the planet.
The study conducted in this paper is 2D; 3D modeling would be required to study the relation between spiral patterns in the vertical component of the velocity and the in-plane radial and azimuthal velocity.
This could give more insight in this particular case, but that is beyond the scope of this paper.

\subsection{IM Lup disk}
\label{ss:imlup}
\begin{figure*}
    \centering
    \includegraphics[width=1\textwidth]{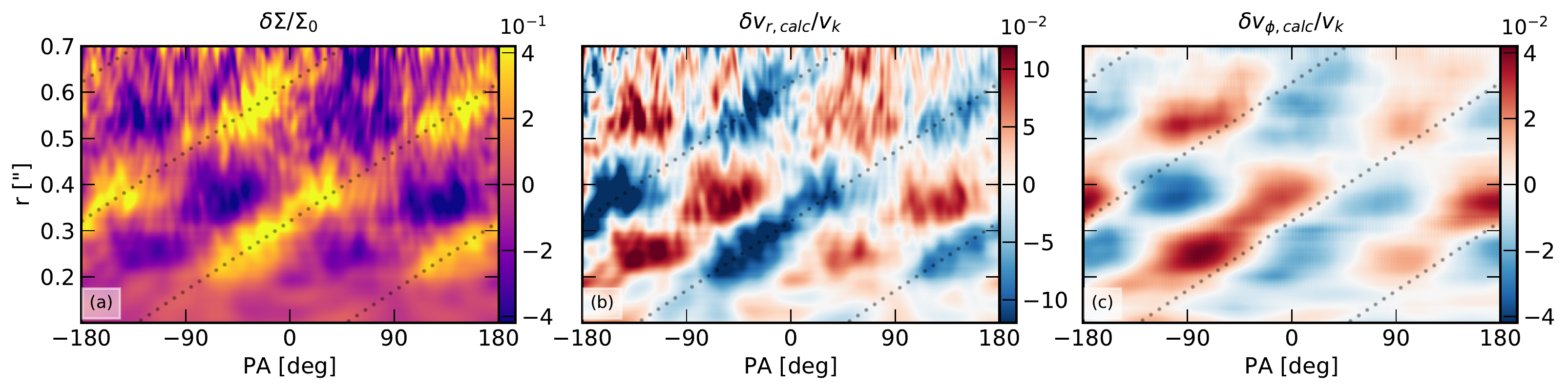}
    \caption{From left to right: the observed surface density perturbation in the millimetre dust continuum of the IM Lup disk normalized to the radial symmetric surface density (a), the perturbation in radial velocity normalized to the Keplerian velocity, derived from the surface density perturbation using Eq. \ref{eq:dens_from_angle} (b) and the azimuthal velocity perturbation normalized to the Keplerian velocity derived from $\delta v_{r,\mathrm{calc}}$ using Eq. \ref{eq:delvr_from_delvphi} (c). The dashed lines represent a fit by eye through the maximum of the spiral in the surface density which serves as a guide for the eye.}
    \label{fig:IMLup_derived_overview}
\end{figure*}
\begin{figure*}
    \centering
    \includegraphics[width=1\textwidth]{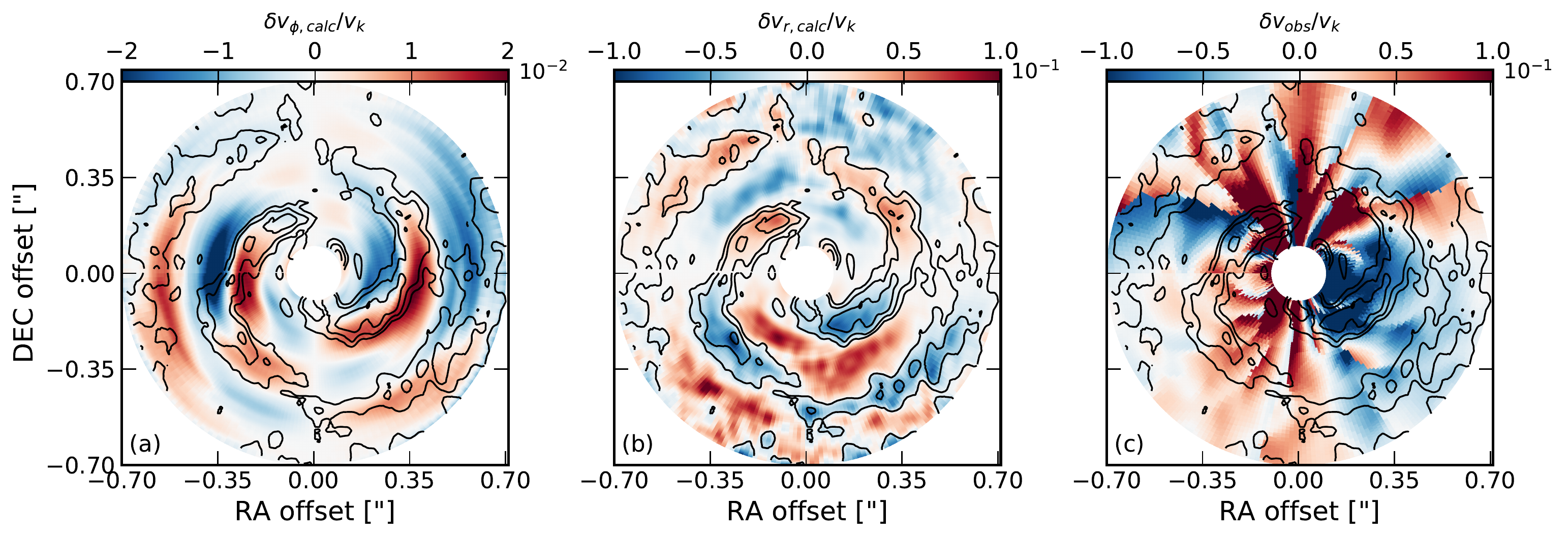}
    \caption{Spirals in gas and dust for the IM Lup disk. From left to right: the estimated perturbation in azimuthal velocity from the dust continuum using Eq. \ref{eq:delvr_from_delvphi} and Eq. \ref{eq:dens_from_angle} (a), the estimated perturbation in radial velocity from the continuum using Eq. \ref{eq:dens_from_angle}(b), the observed velocity residuals in the optically thick CO gas emission after subtracting a Keplerian model of the data (c)}
    \label{fig:IMLup_comparing}
\end{figure*}

We can use the relations derived in Sect. \ref{sec:results_spiral_modelling} on observational data as a prediction of what spiral arms will look like in different velocity components.
As we have shown in Sect. \ref{ss:potential sources}, the best target for this is the IM Lup disk as it has high contrast spiral signal and no cloud contamination.
IM Lup is a 0.5 Myr old K5 star located at a distance of 158 $\pm$ 1 pc \citep{2017A&A...600A..20A,2018yCat.1345....0G}.
The 1.25 mm continuum and $^{12}$CO $J$ = 2-1 gas emission of this disk is observed as part of the DSHARP survey.
A detailed overview of the survey, including the observational setup, calibration and imaging is provided in \citet{2018ApJ...869L..41A}; a detailed description of the spirals in the mm continuum is provided in \citet{2018ApJ...869L..43H}.

After deprojection of the disk using a position angle of 144$^\mathrm{o}$.5 and an inclination of 47$^\mathrm{o}$.5 \citep{2018ApJ...869L..42H} we estimate $\Sigma_\mathrm{0}$ by taking the median value of the surface density along azimuth and the spiral perturbation ($\delta\Sigma$) as the deviation from this value in the data.
In Fig. \ref{fig:IMLup_derived_overview} we show the spiral perturbation in the millimetre dust continuum surface density (left panel) with the derived perturbation in the radial velocity (middle panel) and azimuthal velocity (right panel), derived using Eq. \ref{eq:delvr_from_delvphi} and Eq. \ref{eq:dens_from_angle}.
The pitch angle of the spiral that is indicated in Fig. \ref{fig:IMLup_derived_overview} and used in the model, is determined using a fit by eye through the spiral maximum in the surface density perturbation, assuming that the two parts of the spiral separated by a radial symmetric gap are part of one spiral arm.
Using a parametrization of $\phi$ = $ar + b$ we find $\phi$ = 590[$^\mathrm{o}$]$r$["] - 12[$^\mathrm{o}$] and $\phi$ = 590[$^\mathrm{o}$]$r$["] - 192[$^\mathrm{o}$], which is consistent with the results obtained in \citet{2018ApJ...869L..43H}.
There is evidence of a gap in the continuum emission at a radius of 0.70" \citep{2018ApJ...869L..43H} and a potential kink in the velocity channel maps at the same radius, with a potential planetary origin \citep{2020ApJ...890L...9P}, so we assume that $r$ < $r_{\rm pl}$ in the region we analyze, adding an additional minus sign in Eq. \ref{eq:dens_from_angle}.

The azimuthal velocity perturbation is 2 or 3 times smaller than the perturbation in radial velocity and is shifted $\sim$ 30$^\mathrm{o}$ with respect to the surface density perturbation.
To compare our results with the measured dynamical perturbation we used the $\texttt{bettermoments}$ package described in \citet{2018RNAAS...2..173T} to determine the velocity map of the CO gas emission.
The velocity map is determined using a fit of a quadratic curve to the three pixels closest to the peak intensity, providing a higher spectral precision than the velocity resolution of the data.
The velocity perturbations can be determined from this map by subtracting a Keplerian rotation profile
To do so we fit a Keplerian model to the data using $\texttt{eddy}$ \citep{2019JOSS....4.1220T}, where the line center is given by 
\begin{equation}
    v_\mathrm{k} = \sqrt{\frac{GM_\mathrm{\ast} r^2}{(r^2 + z^2)^{3/2}}},
\end{equation}
where $z$ is the height in the disk using a power law profile with a power law cut off in the outer disk
\begin{equation}
    z(r) = z_\mathrm{0}r^\psi + z_\mathrm{1}r^\varphi.
\end{equation}
As the inclination and the mass of the star are degenerate with each other, we fix the inclination to the same value as found in the continuum.
The resulting parameters of the fit can be found in Table \ref{tab:diskfit_params}.
\begin{table}[!b]
    \centering
    \caption{Parameters of the IM Lup gas-disk fit}
    \label{tab:diskfit_params}
    \begin{tabular}{p{2cm}p{6cm}}\hline\hline
        $x_\mathrm{0}$ & 0.0223"\\
        $y_\mathrm{0}$ & 0.00255"\\
        PA & 143$^\mathrm{o}$.6 \\
        $V_\mathrm{lsr}$ & 4516.5 m/s\\
        $M_\ast$ & 1.275 $M_\mathrm{\oplus}$\\
        $z_\mathrm{0}$ & 0.87\\
        $\psi$ & 1.8\\
        $z_\mathrm{1}$ & -0.59\\
        $\varphi$ & 2.0\\
        $i$ & 47$^\mathrm{o}$.5 (fixed)\\
        dist & 158.4 pc (fixed)\\\hline\hline
    \end{tabular}
\end{table}
In order to compare our computational results with the observations, we project the velocity residuals to what we would be able to see with the known disk inclination.
The radial velocity perturbation is projected as $\delta v_{r\mathrm{,proj}} = \delta v_r$cos(PA + 90$^\mathrm{o}$)sin($i$) and the azimuthal velocity perturbation as $\delta v_{\phi\mathrm{,proj}} = \delta v_\phi$cos(PA)sin($i$) with PA the position angle in the disk and $i$ the inclination.
The results are illustrated in Fig. \ref{fig:IMLup_comparing}, where we show a projected estimate of the azimuthal velocity perturbation derived from the dust continuum emission using Eq. \ref{eq:delvr_from_delvphi} and Eq. \ref{eq:dens_from_angle} (left panel) and an estimate of the radial velocity perturbation using Eq. \ref{eq:dens_from_angle} (central panel).
In the right panel of Fig. \ref{fig:IMLup_comparing} we show the observed velocity residual map.
Our estimate in radial velocity is of the same order of magnitude as the residuals in the observed velocity map. However, the data contain ``spokes'' artifacts that are due to the velocity channelization of the data. With the current velocity resolution, it is therefore impossible to deduce any spiral structures, highlighting the need for data with higher velocity resolution.

\section{Discussion} 
\label{sec:discussion}
We have run $\texttt{FARGO-3D}$ simulations and analyzed the spiral created by the wake of a planet in the different dust components.
We find that dust grains with typical Stokes numbers (e.g., $\mathrm{St} = 0.1$) such that they radially drift fairly quickly \citep[see e.g. ][]{2002ApJ...581.1344T,1972fpp..conf..211W} are closely coupled to the gas in terms of the spiral morphology. This is due to the fact that the azimuthal width of the spiral is much smaller than a whole orbit, which means that the dust grains have to decouple more from the gas in order to react to the perturbation caused by the planet.

\subsection{Importance of $\alpha_\mathrm{\nu}$ and r}
\label{ssec:discussion_dependencealphar}
\noindent\textbf{Viscosity}\\
Throughout the semi-analytical analysis we used a value of $\alpha_\mathrm{\nu} = 10^{-2}$ for the viscosity to test and compare the derived relations.
All relations used are independent of $\alpha_\mathrm{\nu}$, although the shape of the spiral along azimuth changes when using lower values for the viscosity; we chose a relatively high value for $\alpha_\mathrm{\nu}$ to smooth out high frequency oscillations along the position angle.
Changing $\alpha_\mathrm{\nu}$ does not change the pitch angle of the spiral in the gas, but has effect on the amplitude and width of the spiral as the perturbation diffuses out more efficiently with higher viscosity.
This can slightly affect the pitch angle and amplitude of the spiral in the dust components. 
Once normalizing to the properties of the spiral in the gas (see Fig. \ref{fig:ang_and_amp_vs_st}), the shape of the pitch angle and spiral amplitude as function of Stokes number do not change significantly.
\\

\noindent\textbf{Radius}\\
In order to make sure that no boundary effects play a role in our analysis, we analyzed the spiral at $r$ = 1.7 $r_{\rm pl}$ and run the code once with a much larger extent 0.25 $r_{\rm pl}$ < $r$ < 5 $r_{\rm pl}$ (N$_\phi$ x N$_r$ = 1024 x 512), but apart from regions close to the boundary, this does not change the morphology of the spiral in the analyzed part of the disk (0.4 $r_{\rm pl}$ < $r$ < 3 $r_{\rm pl}$).
We analyzed the spiral at a radius of 1.7 $r_{\rm pl}$, far enough from the planet to avoid effects of gap clearing and direct gravitational effects of the planet.
The model is unable to recover the spiral in the dust accurately close to planet, mainly due to nonlinearity of the pitch angle.
The radial dependence of the model is small in the regime where the pitch angle is linear and the critical Stokes number changes only a factor of a few over radius (See Appendix \ref{app:Stc_vs_r}), which finds its main origin in diffusely broadening of the spiral.

\subsection{Small vs large dust grains}
\label{ssec:discussion_hd100453}
\citet{2020MNRAS.491.1335R} showed that the pitch angle of sub-millimetre dust continuum spirals, tracing large dust grains, in HD100453 ($\sim$6$^\mathrm{o}$) are lower than the same spirals in scattered light, tracing small dust grains, ($\sim$16$^\mathrm{o}$).  
Using hydrodynamical simulation they show that this is a general property of spirals as a result of different sound speeds, due to the fact that for externally irradiated disks the midplane is colder than the upper layers of the disk.
This mechanism can create differences in the morphology of observed spirals between gas and dust as well.
One assumption they made in the modeling is that the spirals in the ALMA dust continuum image trace the same morphology as the spirals in the midplane gas.
In a typical disk, the critical Stokes number when gas and dust spirals start to deviate, determined in Sect. \ref{sssec:spirals_vs_ST}, corresponds to centimeter sized dust grains, well above the maximum grain size that can be reached due to the bouncing and fragmentation barriers \citep[e.g.,][]{2010A&A...513A..79B,2012A&A...540A..73W}.
This means that the assumption holds that the spirals in ALMA dust continuum trace the same morphology as the midplane gas.

\subsection{Spiral origin}
\label{ssec:discussion_spiral_origin}
Spiral perturbations in protoplanetary disks can be created in two different ways.
In this paper we focused on spirals that are created by Lindblad resonances in the disk, generated by planets.
This differs significantly from spirals caused by gravitational instability of massive disks, as these instabilities are thought to move with the flow. This means that self-gravitating spirals can trap dust \citep[e.g.,][]{2004MNRAS.355..543R,2006MNRAS.372L...9R,2016MNRAS.458.2676B} and their amplitude \textit{increases} with Stokes number. 
Instead, in the planet case spirals move at the same angular velocity as the planet throughout the disk; spiral crossing is short (compared to the orbital time-scale) and there is no particle trapping. 
In this case, the amplitude of the spiral \textit{decreases} with Stokes number, as we have shown in this paper.
In principle, this difference can be used in observations to discriminate between the two origins using continuum multi-wavelength observations.

In the IM Lup system there is evidence of a gap in the continuum emission at a radius of 0.70" or 111 au \citep{2018ApJ...869L..43H} and a potential kink in the velocity channel maps at the same radius with a potential planetary origin \citep{2020ApJ...890L...9P}, which makes the assumption that the spiral is planet-launched reasonable.
$M_\mathrm{disk}/M_\mathrm{\ast}$ is considered too small to cause gravitational instability \citep[e.g.][]{2016ARA&A..54..271K}, although there is some debate on the origin of the spiral \citep{2018ApJ...869L..43H}.
If the spirals we observe in the continuum of IM Lup are generated by gravitational instability, our model would not work, and the estimate of the projected dynamics derived from the surface density would be incorrect.
In addition to differentiate between grains of different sizes, data with a better velocity resolution can be used to end this debate, by comparing the velocity perturbations from Keplerian rotation with our estimates based on the planetary origin.

\subsection{Limitations and future prospects}
In this paper we did not consider the impact of dust evolution on the local dust density structures. Dust grain coagulation and fragmentation can potentially have a strong impact on the local dust density structures as well \citep[see e.g.][]{2019ApJ...885...91D}.
However, in contrast to for example gaps and rings, the spiral structure is established in only a few orbits, while grain growth is a significantly slower process: measured in orbits, the collisional timescale between two grains is of order of the gas-to-dust ratio (e.g., \citealt{2012A&A...539A.148B}). It is thus a reasonable assumption to take the grain properties as fixed as we have done in this paper. This does not exclude the possibility that, on secular timescales, grain growth will modify the grain properties and the spiral will readjust accordingly, following the relations we have derived in this work.
The same argument holds for another effect we neglected in this paper, namely planet migration. The spiral reaches steady-state in a few orbits, which is fast compared to the migration time scale (hundreds or more orbits, e.g. \citealt{Baruteau2014}). This justifies the assumption made in Sect. \ref{sec:results_spiral_modelling} that the planet moves on a fixed circular orbit.

Another effect we neglected in this paper is the feedback of the dust onto the gas. This would not be justified close to the pressure maximum created by the planet, where the dust accumulates and dust-to-gas ratio becomes high. However, in our analysis we focused on the spiral structure further from the planet, where there is no accumulation effect. It should be noted that recently \citet{2018MNRAS.479.4187D} pointed out that the effect of dust feedback could be non-negligible even in smooth parts of the disk, away from pressure maxima, with the magnitude of the effect depending on the value of $\alpha_\mathrm{\nu}$ (though see e.g. \citealt{2019MNRAS.486.4829R} for calculations including the effect of dust feedback where this is not seen). If this is confirmed in future studies, the effect of dust feedback on planetary spirals would need to be re-assessed. Our paper provides the starting point to build a theoretical description including dust feedback.

Since the $^{12}$CO line is optically thick, it is useful to observe the dynamics of rarer isotopologues to study the in-plane motions in the spiral, as the in-plane perturbation fades out higher up in the disk.
However, studying the dynamics of for example C$^{18}$O does take a considerable amount of observing time.
Understanding velocity perturbations in the vertical direction can be useful to get a better understanding of the mechanism that drives the spiral in the gas, for example in TW Hya.
3D simulations and analysis needs to be done in order to account for vertical motions in our model, but that is beyond the scope of this paper.

For now, it is not possible to determine the gas surface density perturbation of observations based on the continuum and vice versa, since the velocity of the spiral with respect to the disk is required (i.e. the position of the planet).
Since we show in Fig. \ref{fig:ang_and_amp_vs_st} that the difference in spiral morphology and amplitude changes only slightly for millimeter emission, we used the approximation that the spiral is unaffected in the continuum emission in order to test our model and make a prediction for IM Lup.
If in the future spirals in both continuum and gas components will be observed, this property might be useful to determine the location of the planet that launches the spiral.

\section{Conclusions}
\label{sec:conclusions}
In this paper, we systematically studied the variations in the properties of planet-induced spirals in gas and dust caused by the dynamics. We can conclude the following:
\begin{itemize}
    \item The morphology and amplitude of a planet-driven spiral only changes for very large dust grains. The shape of the curve amplitude as function of Stokes number is almost independent of planet mass, making it suitable as a tracer of dust grain size if a spiral signal in both gas and continuum is detected. 
    \item A planet-launched spiral in surface density finds its origin in a perturbation in radial velocity, which can in turn be described as the immediate result of a perturbation in azimuthal velocity loosing or gaining angular momentum, hence pushing matter in or out.
    \item The azimuthal offset and amplitude change between spirals in gas and dust can be explained using the fact that dust grains feel the drag of the gas, but do not react fast enough, causing the curve to flatten.
    \item The above two points can be combined in a model that accurately estimates the morphology and amplitude in surface density of the spiral in dust with St $\leq$ 1, using only the perturbation in azimuthal velocity of the gas.
    \item We find a new, convincing spiral in the dynamics of the gas in the TW Hya disk with a kink at the position of the gap in the continuum that matches well with the theoretical shape of a spiral wake. This spiral extends along both the major and minor axis of the disk, which means that we see a spiral pattern of gas that is launched from the disk in vertical direction, probably driven by a planet potential.
    \item We prove that our model is able to handle observational data and can be used to make an estimate of how spirals will look like in in-plane gas dynamics. However, data with better velocity resolution are needed to be able to compare the continuum spirals with spirals in the gas dynamics.
\end{itemize}

\begin{acknowledgements}
We thank the referee for their valuable comments. We also thank Ewine van Dishoeck for useful discussions as well as a careful reading of our manuscript and many constructive comments, Myriam Benisty for invaluable discussions and Richard Teague for an effortless providing of the reduced data of the TW Hya disk. G.R. acknowledges support from the Netherlands Organization for Scientific Research (NWO, program number 016.Veni.192.233).
\end{acknowledgements}

\bibliographystyle{AA/aa}
\bibliography{paper}

\begin{appendix}
\section{Interpretation of line density and Eq. 14}\label{app:interpretation_linedensity}
\begin{figure}[ht]
    \centering
    \includegraphics[width=1\linewidth]{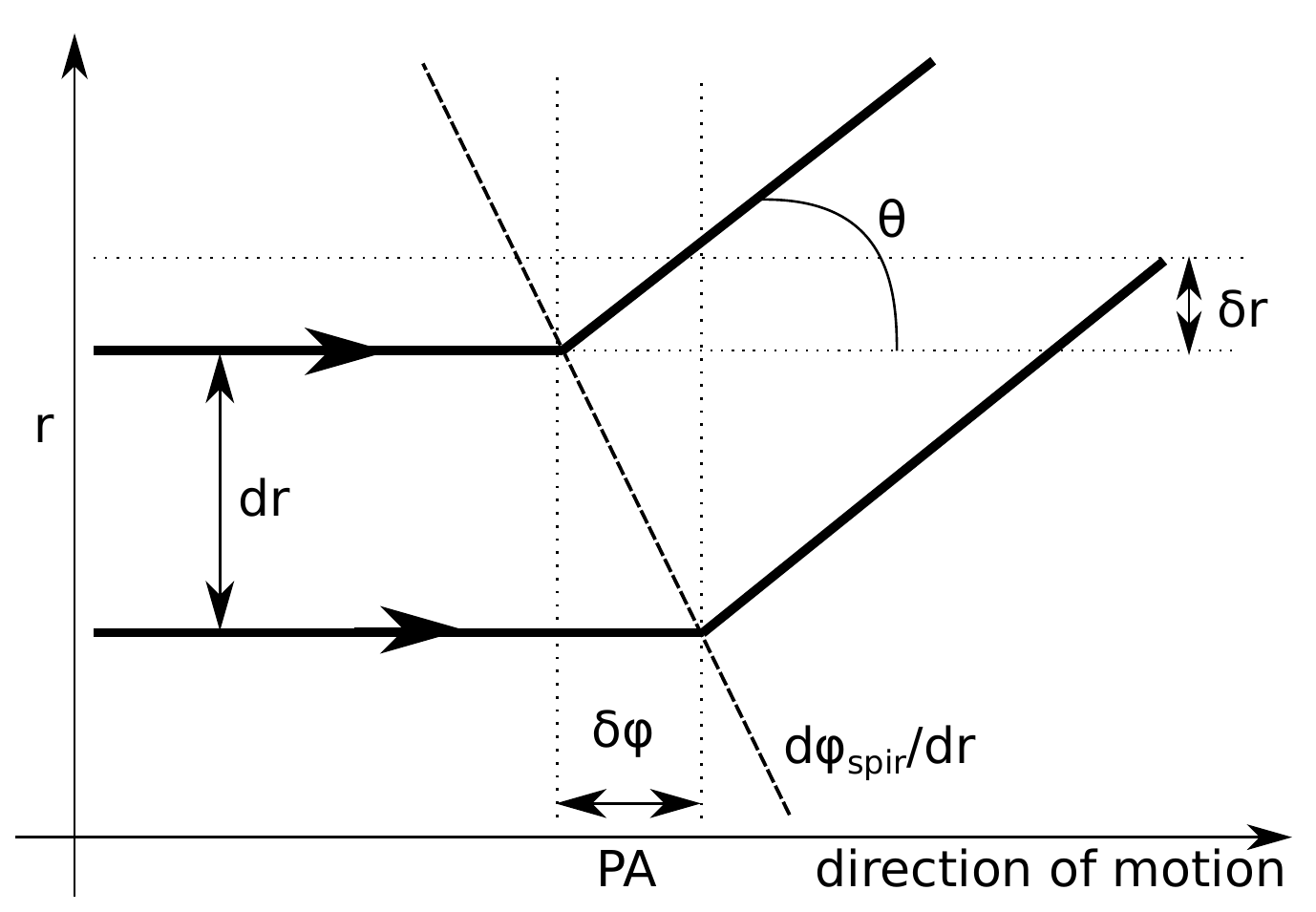}
    \caption{A toy model of an infinitesimal small part of the stream lines a particle will follow during its orbit around the central object. d$r$ is the median distance between the stream lines, $\delta r$ is the radial perturbation, $\delta\phi$ is the azimuthal shift between the stream lines, caused by the pitch angle of the spiral ($\mathrm{d}\phi_\mathrm{{spir}}/\mathrm{d}r$) and $\theta$ is the angle of the stream line from circular orbit.}
    \label{fig:geometry_density}
\end{figure}
Figure \ref{fig:geometry_density} shows a toy model of an infinitesimal small part of the stream lines that particles will follow during an orbit around the central object, separated from each other by an infinitesimally small change in orbit radius, d$r$.
The azimuthal shape of the density perturbation for these two stream lines is the same and the pitch angle of the spiral causes a small azimuthal shift between the streamline profiles, $\delta\phi$.
This azimuthal shift results in a change in radial distance between the streamlines $\delta r$ at certain position angles.

The result of Eq. \ref{eq:dens_from_angle} can be checked using this simple geometrical model. Defining a line density as the number of stream lines in an arbitrary radial interval, we can relate the normalized perturbation in line density (-$\delta r$/d$r$, shown in red in Fig. \ref{fig:spiral_dynamics}) to the normalized perturbation in surface density ($\delta \Sigma/\Sigma_\mathrm{0}$):
\begin{equation}
    \frac{\delta\Sigma}{\Sigma_\mathrm{0}} = -\frac{\delta r}{\mathrm{d}r}.
\end{equation}

We can find the normalized line density perturbation ($\delta r$/d$r$) by using that the pitch angle of the spiral fixes d$r$/$\delta\phi$ = $d\phi_\mathrm{spir}/\mathrm{d}r$ and that the angle $\theta$, as indicated in Fig. \ref{fig:geometry_density}, is given by tan($\theta$) = $\delta r/\delta \phi$. Since $\delta r$ can be written in terms of the radial velocity as $\delta r = v_r\mathrm{d}t$ and $\delta \phi$ can be written in terms of the angular velocity as $\delta \phi$ = $\Omega_\mathrm{k}\mathrm{d}t$ we can write this as tan($\theta$) = $\delta r/\delta \phi$ =  $v_r/\Omega_\mathrm{k} = rv_r$/$v_\phi$. 
Dividing these two quantities gives:
\begin{equation}
    \delta\Sigma = -\Sigma_\mathrm{0}r\frac{\delta v_r}{v_{\phi\mathrm{,0}}}\frac{\mathrm{d}\phi(r)_\mathrm{sp}}{\mathrm{d}r},
\end{equation}
which is the exact same equation as Eq. \ref{eq:dens_from_angle}.

\newpage
\section{Result of the different substeps}\label{app:substeps}
\begin{figure}[ht]
    \centering
    \includegraphics[width=1\linewidth]{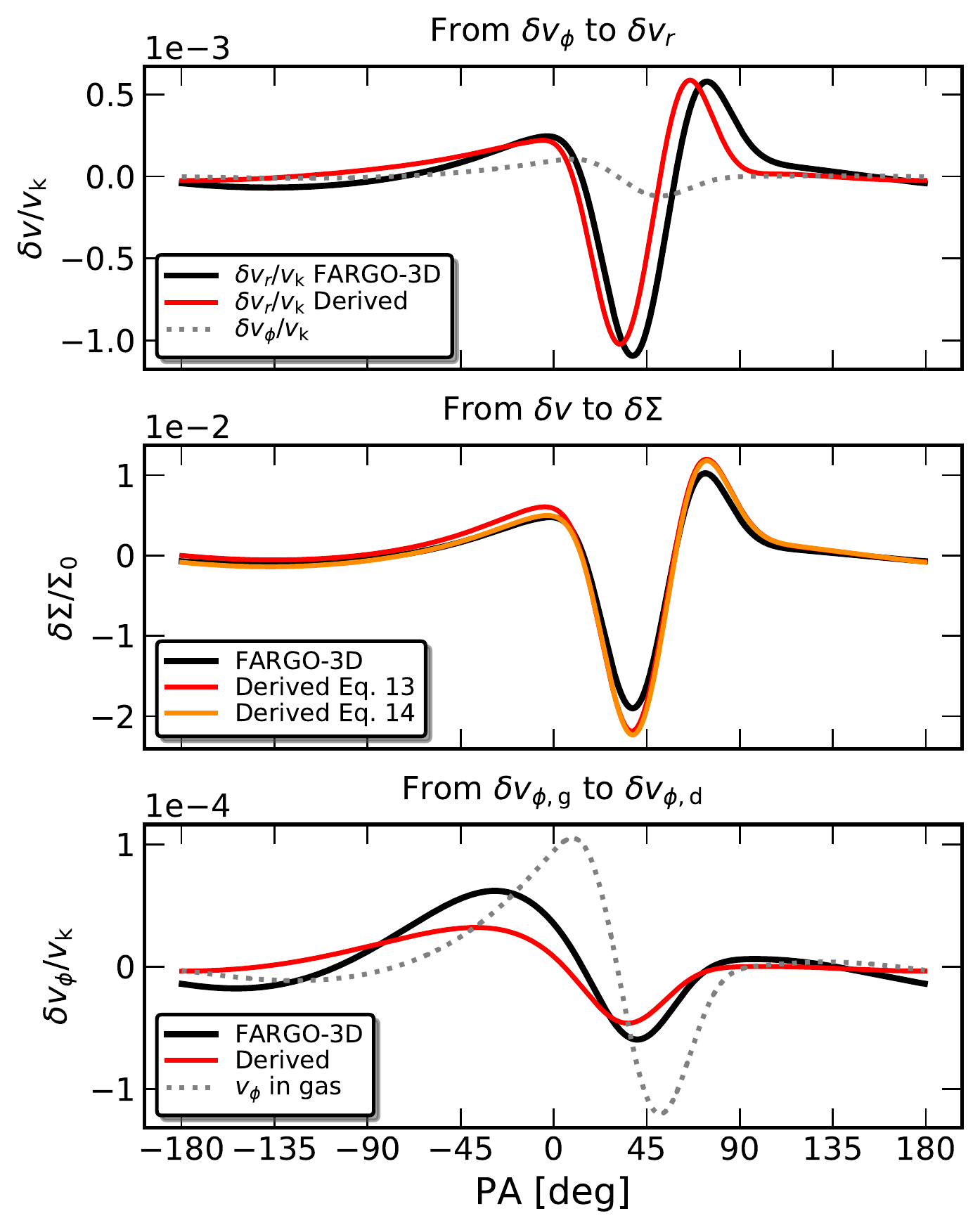}
    \caption{Results of the different substeps as discussed in Sect. \ref{ssec:results_combining_everything}. Top panel: The perturbation in radial velocity as determined by $\texttt{FARGO-3D}$ and determined using Eq. \ref{eq:delvr_from_delvphi} with the azimuthal velocity perturbation as a reference. Central panel: The perturbation in surface density as determined by $\texttt{FARGO-3D}$ and determined from the perturbation in velocity using Eq. \ref{eq:dens_from_angle} and Eq. \ref{eq:delsigma_from_delvr}. Bottom panel: result of the numerical integration to derive the azimuthal velocity perturbation of the dust from the azimuthal velocity of the gas, together with the $\texttt{FARGO-3D}$ results and the azimuthal velocity perturbation as comparison.}
    \label{fig:result_dif_steps}
\end{figure}

\newpage
\section{St$_\mathrm{c}$ as function of $r$}
\label{app:Stc_vs_r}
\begin{figure}[ht]
    \centering
    \includegraphics[width=1\linewidth]{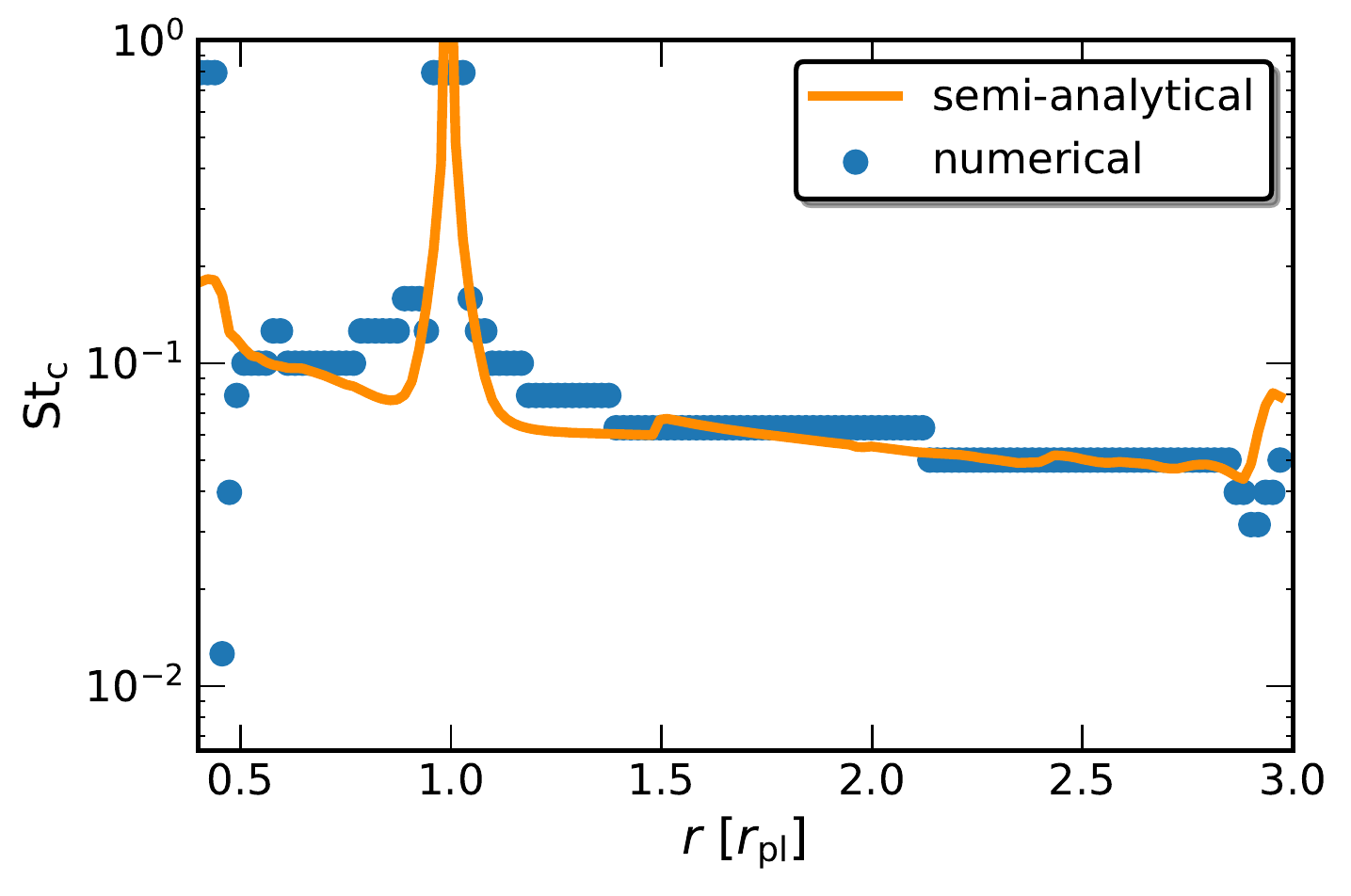}
    \caption{Critical Stokes number as function of radius. Note that the changes in St$_\mathrm{c}$ are minor, apart from positions near the planet or boundary. The semi-analytical approach follows the numerical approach accurately, which means that the differences are mostly caused by a change in spiral width.}
    \label{fig:Stc_vs_r}
\end{figure}

\end{appendix}
\end{document}